\documentclass[onecolumn]{mn2e}
\newif\ifAMStwofonts

\def\gtsim {>\kern-1.2em\lower1.1ex\hbox{$\sim$}}
\def\ltsim {<\kern-1.2em\lower1.1ex\hbox{$\sim$}}
\def\gtsim {>\kern-1.2em\lower1.1ex\hbox{$\sim$}}
\def\ltsim {<\kern-1.2em\lower1.1ex\hbox{$\sim$}}
\def\ref{\hangindent=1pc \hangafter=1 \noindent}
\usepackage{epsfig}
\begin{document}

\title[$R$-modes of a magnetized neutron star]
{$R$-modes of a neutron star with a magnetic dipole field}

\author[U. Lee]{Umin Lee$^1$\thanks{E-mail: lee@astr.tohoku.ac.jp}
\\$^1$Astronomical Institute, Tohoku University, Sendai, Miyagi 980-8578, Japan}

\date{Typeset \today ; Received / Accepted}
\maketitle


\begin{abstract} 
We study $r$-modes of a rotating magnetized neutron star,
assuming a magnetic dipole field whose axis is aligned with the axis of rotation.
We approach the problem by applying a singular perturbation theory to the oscillations
of rotating stars.
In this treatment, we divide the star 
into a thin surface magnetic layer and a non-magnetic core.
We integrate linearized ideal MHD equations in the surface magnetic layer and 
non-magnetic oscillation equations in the core, and match the two integrations at the interface
to obtain a complete solution.
For a polytropic neutron star model of mass $M=1.4M_\odot$ and radius $R=10^6$cm, 
the magnetic dipole field becomes effective on
the modal properties of the $r$-modes
only when the field strength $B_S$ is much greater than $10^{14}$G.
We also find that
the damping effects caused by very short magnetic perturbations in the surface layer
are not important for the $r$-mode instability of rapidly rotating neutron stars if
the field strength $B_S$ is smaller than $10^{12}$G.
\end{abstract}

\section{Introduction}

Several numerical investigations have recently been carried out to study the properties of
acoustic modes of magnetized stars
(e.g., Dziembowski \& Goode 1996, Bigot et al 2000, Cunha \& Gough 2000,
Saio \& Gautschy 2004), particularly for roAp stars (Kurtz 1990).
These authors solved the problem of the oscillations of a magnetized star 
by applying a singular perturbation theory to the stellar oscillation.
In their treatment, the stellar interior was divided
into the thin surface magnetic layer where the magnetic pressure $p_m=B^2/8\pi$
is comparable to or greater than the gas pressure $p$ and
the non-magnetic core where the gas pressure is dominant over the magnetic pressure
(Biront et al 1982, Roberts \& Soward 1983, Cambell \& Papaloizou 1986).
Linearized ideal MHD equations were integrated in the thin surface magnetic layer, 
and non-magnetic oscillation equations were integrated in the core, and then
the two integrations were matched at the 
interface between the magnetic and non-magnetic regions to make a complete solution.
This treatment is based on the following properties of magnetic perturbations in the interior of stars.
If we assume that the gass pressure dominates the magnetic pressure, or equivalently
that the Alfv\'en wave velocity $v_A=\sqrt{B^2/4\pi\rho}$ 
is much slower than the sound velocity $v_S\sim \sqrt{p/\rho}$ in the deep interior,
it is reasonable to consider that the magnetic perturbations are largely determined by the induction equation.
But, there also exists a component of magnetic perturbations that originates from 
the Lorentz force term $\bmath{j}\times \bmath{B}$ in the equation of motion.
For a typical oscillation frequency of acoustic modes, 
this component appears as very short magnetic perturbations, the wavelengths of which are
much shorter than those of the acoustic modes.
In the limit of $\beta\equiv p/p_m\rightarrow\infty$ in the deep interior, 
we can expect the short wave component of
the magnetic perturbations may be lost from the wave system suffering strong dissipation,
which makes it legitimate to divide the stellar interior into a magnetic surface layer and a non-magnetic core
and to disregard the short wave component in the core
(e.g., Biront et al 1982, Roberts \& Soward 1983).
If we assume that
the oscillation energy imparted by the short magnetic perturbations is lost from the wave system
(Biront et al 1982, Roberts \& Soward 1983), this $loss$ of the magnetic perturbation energies works as a
damping mechanism for the acoustic modes.
This damping mechanism was implemented in global acoustic mode calculations, for example, by 
Bigot et al (2000), Cunha \& Gough (2000), Saio \& Gautschy (2004).

Although the studies mentioned above were all for non-rotating stars, 
Shibahashi \& Takata (1993) and Takata \& Shibahashi (1995) have developed a theory of 
the oscillations of rotating magnetized stars, treating both rotation and the magnetic field 
as small perturbations to the oscillations of non-rotating and non-magnetic stars. 
They recognized, however, that the perturbational treatment of the magnetic field in the surface layer
might not be justified.
Bigot \& Dziembowski (2002) have also considered the effects of a magnetic field and rotation
on acoustic oscillations.
Although they assumed a slow rotation to treat high radial order $p$-modes, 
they employed a non-perturbative approach for the magnetic field, to improve the analyses
by Shibahashi \& Takata (1993) and Dziembowski \& Goode (1996).

$R$-modes of neutron stars have attracted good attention recently because of the 
recognition that the $r$-modes are driven unstable by the emission of gravitational radiation
(Andersson 1998, Friedman \& Morsink 1998) 
due to Chandrasekhar-Friedman-Schutz instability mechanism 
(Chandrasekhar 1970, Friedman \& Schutz 1978 a,b).
The $r$-modes of rotating neutron stars are expected to be a good candidate for
a gravitational wave source (e.g., Andersson \& Kokkotas 2004).
They are also regarded as a mechanism that decelerates the spin of 
young neutron stars, since the gravitational radiation carries away angular momentum
of the star (e.g., Owen et al 1998).
However, before we draw a decisive conclusion concerning the role played by the $r$-modes
in neutron star physics, we need to examine how strongly the modal properties are
affected by various physical factors such as
viscosity, solid crust, superfluidity, general relativity, 
nonlinear amplitude saturation, and so on.
A magnetic field of the star is one of such influential factors.

$R$-mode oscillations of rotating magnetized neutron stars have been discussed by  
Rezzolla, Lamb, \& Shapiro (2000), Ho \& Lai (2000), and  Morsink \& Rezania (2002).
For example, Ho \& Lai (2000) estimated the shifts of the $r$-mode frequency
caused by the strong magnetic field
using a local linear analysis, suggesting that the oscillation manifests itself as Rossby
waves in the bulk interior, but assumes the Alfv\'en wave character near the stellar surface.
Morsink \& Rezania (2002) developed, based on the formalism introduced by Schenk et al (2002),
a formalism to compute the magnetically modified global $r$-modes, without assuming a
weak magnetic field of the star.
In this paper, we approach the problem of the oscillations of a rotating magnetized star by
applying a singular perturbation theory to 
the oscillations of a rotating non-magnetic star (e.g., Lee \& Saio 1986).
In \S 2, we derive the oscillation equations of a rotating magnetized star, assuming
the axis of the dipole field is aligned with the axis of rotation.
\S 3 is for numerical results, and \S 4 is for discussions and conclusions.

\section{method of solution}

\subsection{Oscillation equations of a rotating star with a magnetic dipole field}

In this paper, we assume adiabatic oscillations of a uniformly rotating star
and employ the Cowling
approximation, neglecting the Euler perturbation of the gravitational potential.
We also assume that the axis of the magnetic dipole field is aligned with that of rotation.
If we employ spherical polar coordinates $(r,\theta,\phi)$ whose origin is at the center
of the star, a magnetic dipole field may be given by
\begin{equation}
\bmath{B}=\mu\nabla\left(\cos\theta/r^2\right),
\end{equation}
where $\mu$ is the magnetic dipole moment, and we have assumed that the axis of the field corresponds to
the coordinate axis defined by $\theta=0$.
When the axes of rotation and the magnetic field are aligned, 
it is possible to assume that the equilibrium state is axisymmetric with respect to
the axe $\theta=0$, and that the time and azimuthal angular dependence of the perturbations
is given by $e^{i\sigma t+im\phi}$, where $\sigma$ is the
oscillation frequency in an inertial frame and $m$ is an integer representing the azimuthal 
wave number.
Note that the magnetic dipole field is a force free field 
(i.e., $(\nabla\times \bmath{B})\times \bmath{B}=0$) and does not affect
the equilibrium structure of the star.
Since we are interested in low frequency oscillations in this paper, we neglect 
the effects of rotational deformation.
Linearized ideal MHD equations with infinite conductivity may be given by
(e.g., Unno et al 1989)
\begin{equation}
-\omega^2\bmath{\xi}=-{1\over\rho}\nabla p^\prime+{\rho^\prime\over\rho^2}\nabla p
-2i\omega\bmath{\Omega}\times\bmath{\xi}
+{1\over4\pi\rho}(\nabla\times\bmath{B}^\prime)\times\bmath{B},
\end{equation}
\begin{equation}
\rho^\prime+\nabla\cdot(\rho\bmath{\xi})=0,
\end{equation}
\begin{equation}
\bmath{B}^\prime=\nabla\times(\bmath{\xi}\times\bmath{B}),
\end{equation}
and for adiabatic pulsations
\begin{equation}
{\rho^\prime\over\rho}={1\over\Gamma_1}{p^\prime\over p}-\xi_r A,
\end{equation}
where $\rho$ is the mass density, $p$ is the pressure,
$\bmath{\Omega}$ is the angular velocity of rotation, $\bmath{B}$ is the magnetic field,
and the physical quantities with a prime $(^\prime)$ denote their Euler perturbations, 
$\bmath{\xi}$ is the displacement vector, and
$\omega=\sigma+m\Omega$ denotes the oscillation frequency observed in the corotating frame
of the star.
Here, $A$ is the Schwartzshild discriminant defined by
\begin{equation}
A={d\ln\rho\over dr}-{1\over\Gamma_1}{d\ln p\over dr},
\end{equation}
where
\begin{equation}
\Gamma_1=\left({\partial\ln p\over\partial\ln \rho}\right)_{ad}.
\end{equation}

To represent the angular dependence of the perturbations of a rotating magnetized star, we expand 
the perturbed quantities in terms of spherical harmonic functions 
with different degrees $l$ for a given $m$.
The displacement vector $\bmath{\xi}$ is then given by
\begin{equation}
\xi_r=\sum_{j=1}^{\infty}S_{l_j}(r)Y_{l_j}^m(\theta,\phi)e^{i\sigma t},
\end{equation}
\begin{equation}
\xi_\theta=\sum_{j=1}^{\infty}
\left[H_{l_j}(r){\partial\over\partial\theta}Y_{l_j}^m(\theta,\phi)
+T_{l^\prime_j}(r){1\over\sin\theta}{\partial\over\partial\phi}Y_{l^\prime_j}^m(\theta,\phi)\right]
e^{i\sigma t},
\end{equation}
\begin{equation}
\xi_\phi=\sum_{j=1}^{\infty}
\left[H_{l_j}(r){1\over\sin\theta}{\partial\over\partial\phi}Y_{l_j}^m(\theta,\phi)
-T_{l^\prime_j}(r){\partial\over\partial\theta}Y_{l^\prime_j}^m(\theta,\phi)\right]
e^{i\sigma t},
\end{equation}
and the perturbed magnetic field $\bmath{B}^\prime$ is given by
\begin{equation}
B^\prime_r=B_0(r)\sum_{j=1}^{\infty}
b^S_{l^\prime_j}(r)Y_{l^\prime_j}^m(\theta,\phi)e^{i\sigma t},
\end{equation}
\begin{equation}
B^\prime_\theta=B_0(r)\sum_{j=1}^{\infty}
\left[b^H_{l^\prime_j}(r){\partial\over\partial\theta}Y_{l^\prime_j}^m(\theta,\phi)
+b^T_{l_j}(r){1\over\sin\theta}{\partial\over\partial\phi}Y_{l_j}^m(\theta,\phi)\right]
e^{i\sigma t},
\end{equation}
\begin{equation}
B^\prime_\phi=B_0(r)\sum_{j=1}^{\infty}
\left[b^H_{l^\prime_j}(r){1\over\sin\theta}{\partial\over\partial\phi}Y_{{l^\prime_j}}^m(\theta,\phi)
-b^T_{l_j}(r){\partial\over\partial\theta}Y_{l_j}^m(\theta,\phi)\right]
e^{i\sigma t},
\end{equation}
and a perturbed scalar quantity such as $p^\prime$ is given by
\begin{equation}
p^\prime=\sum_{j=1}^{\infty}p^\prime_{l_j}(r)Y_{l_j}^m(\theta,\phi)e^{i\sigma t}.
\end{equation}
where $B_0(r)=\mu/r^3$,
and $l_j=|m|+2(j-1)$ and $l^\prime_j=l_j+1$ for even modes, and 
$l_j=|m|+2j-1$ and $l^\prime_j=l_j-1$ for odd modes, respectively, and $j=1,~2,~3,~\cdots$.

Substituting the expansions into the linearized basic equations (2) to (5), 
we obtain a set of 
coupled ordinary linear differential equations of infinite dimension
for the expansion coefficients.
If we use for the expansion coefficients vector notation given by
\begin{equation}
(\bmath{y}_1)_j=S_{l_j}/r, \quad (\bmath{y}_2)_j=p^\prime_{l_j}/\rho g r, \quad
(\bmath{b}^S)_j=b^S_{l^\prime_j}, \quad (\bmath{b}^H)_j=b^H_{l^\prime_j}, \quad
(\bmath{b}^T)_j=b^T_{l_j}, \quad (\bmath{h})_j=H_{l_j}/r, \quad
(\bmath{t})_j=T_{l^\prime_j}/r,
\end{equation}
the oscillation equations of a rotating star with an aligned magnetic dipole field 
are given in the Cowling approximation as
\begin{eqnarray}
&&r{d\bmath{y}_1\over dr}=\left(V_G-3\right)\bmath{y}_1-V_G\bmath{y}_2
+\Lambda_0\bmath{h},
\end{eqnarray}
\begin{eqnarray}
&&r{d\bmath{y}_2\over dr}=\left(c_1\bar\omega^2+rA\right)\bmath{y}_1
+\left(1-rA-U\right)\bmath{y}_2-2mc_1\bar\omega\bar\Omega\bmath{h}
-2c_1\bar\omega\bar\Omega C_0i\bmath{t} \nonumber \\
&&+{c_1\bar\omega^2\over 2}\left(C_0r{d\over dr}{\bmath{b}^H\over\alpha}
+mr{d\over dr}{i\bmath{b}^T\over\alpha}
-f C_0{\bmath{b}^H\over\alpha}-mf{i\bmath{b}^T\over\alpha}
-C_0{\bmath{b}^S\over\alpha}\right),
\end{eqnarray}
\begin{eqnarray}
&&r{d\over dr}\left(\matrix{\bmath{b}^H/\alpha \cr i\bmath{b}^T/\alpha\cr}\right)
=\left(\matrix{m\nu G_1^{-1}(K+M_0)-\Lambda_1K/\alpha & G_1^{-1}M_0\Lambda_0/c_1\bar\omega^2\cr
-\nu G_0^{-1}(m^2\Lambda_0^{-1}+M_1\Lambda_1K)&-mG_0^{-1}/c_1\bar\omega^2\cr}\right)
\left(\matrix{\bmath{y}_1\cr\bmath{y}_2\cr}\right) \nonumber \\
&&+\left(\matrix{f\bmath{1}&-mG_1^{-1}(C_1+M_0\Lambda_0)/2\cr
0&f\bmath{1}+G_0^{-1}(m^2\bmath{1}+M_1\Lambda_1C_1)/2\cr}\right)
\left(\matrix{\bmath{b}^H/\alpha \cr i\bmath{b}^T/\alpha\cr}\right)
+\left(\matrix{-G_1^{-1}M_0\Lambda_0-2\Lambda_1M_0/\alpha &
mG_1^{-1}-(\nu+2m/\alpha)\bmath{1}\cr mG_0^{-1}-\nu\bmath{1}&-G_0^{-1}M_1\Lambda_1\cr}\right)
\left(\matrix{\bmath{h}\cr i\bmath{t}\cr}\right),
\end{eqnarray}
\begin{eqnarray}
&&r{d\over dr}\left(\matrix{\bmath{h}\cr i\bmath{t}\cr}\right)=
{1\over2}\left(\matrix{(V_G-4)G_0^{-1}(m^2\Lambda_0^{-1}+M_1\Lambda_1K) &
-V_GG_0^{-1}(m^2\Lambda_0^{-1}+M_1\Lambda_1K)\cr -m(V_G-4)G_1^{-1}(K+M_0)&
mV_GG_1^{-1}(K+M_0)\cr}\right)\left(\matrix{\bmath{y}_1\cr\bmath{y}_2\cr}\right)\nonumber \\ 
&&+{\alpha\over 2}\left(\matrix{G_0^{-1}M_1\Lambda_1&-mG_0^{-1}\cr
-mG_1^{-1}&G_1^{-1}M_0\Lambda_0\cr}\right)
\left(\matrix{\bmath{b}^H/\alpha \cr i\bmath{b}^T/\alpha\cr}\right)
+\left(\matrix{\bmath{1}+G_0^{-1}M_1\Lambda_1C_1/2&-mG_0^{-1}C_0/2\cr
-mG_1^{-1}C_1/2&\bmath{1}+G_1^{-1}M_0\Lambda_0C_0/2\cr}\right)
\left(\matrix{\bmath{h}\cr i\bmath{t}\cr}\right),
\end{eqnarray}
\begin{eqnarray}
&&\bmath{b}^S=-\Lambda_1K\bmath{y}_1-2\Lambda_1 M_0\bmath{h}-2mi\bmath{t},
\end{eqnarray}
where $\nu=2\Omega/\omega$, and
$\bar\omega=\omega/\sqrt{GM/R^3}$ and $\bar\Omega=\Omega/\sqrt{GM/R^3}$ with
$M$ and $R$ being the mass and the radius of the star, and
\begin{equation}
U={d\ln M_r\over d\ln r}, \quad V=-{d\ln p\over d\ln r}, \quad V_G={V\over\Gamma_1}, \quad
c_1={(r/R)^3\over M_r/M}, 
\end{equation}
and
\begin{equation}
\alpha={c_1\bar\omega^2\beta V\over 4}, \quad \beta={p\over B_0^2/8\pi},
\quad f=2-{d\ln\alpha\over d\ln r}.
\end{equation}
Here, equation (16) comes from the continuity equation (3), equations (17) and (18) 
from the radial and the horizontal components of
the equation of motion (2), and equations (19) and (20) from the induction equation (4).
The matrices $G_0$ and $G_1$ are defined as
\begin{equation}
G_0=m^2\Lambda_0^{-1}-M_1\Lambda_1M_0, \quad
G_1=m^2\Lambda_1^{-1}-M_0\Lambda_0M_1,
\end{equation}
and the non-zero elements of the matrices $M_0$, $M_1$, $C_0$, and $C_1$ are given by
\begin{equation}
(M_0)_{j,j}={l_j\over l_j+1}J^m_{l_j+1}, \quad
(M_0)_{j,j+1}={l_j+3\over l_j+2}J^m_{l_j+2}, \quad
(M_1)_{j,j}={l_j+2\over l_j+1}J^m_{l_j+1}, \quad
(M_1)_{j+1,j}={l_j+1\over l_j+2}J^m_{l_j+2},
\end{equation}
\begin{equation}
(C_0)_{j,j}=-(l_j+2)J^m_{l_j+1}, \quad (C_0)_{j+1,j}=(l_j+1)J^m_{l_j+2}, \quad
(C_1)_{j,j}=l_jJ^m_{l_j+1}, \quad (C_1)_{j,j+1}=-(l_j+3)J^m_{l_j+2},
\end{equation}
for even modes, and
\begin{equation}
(M_0)_{j,j}={l_j+1\over l_j}J^m_{l_j}, \quad
(M_0)_{j+1,j}={l_j\over l_j+1}J^m_{l_j+1}, \quad
(M_1)_{j,j}={l_j-1\over l_j}J^m_{l_j}, \quad
(M_1)_{j,j+1}={l_j+2\over l_j+1}J^m_{l_j+1},
\end{equation}
\begin{equation}
(C_0)_{j,j}=(l_j-1)J^m_{l_j}, \quad (C_0)_{j,j+1}=-(l_j+2)J^m_{l_j+1}, \quad
(C_1)_{j,j}=-(l_j+1)J^m_{l_j}, \quad (C_1)_{j+1,j}=l_jJ^m_{l_j+1},
\end{equation}
for odd modes, and $\bmath{1}$ denotes the unit matrix.
The definition of the matrices $K$, $L_0$, $L_1$, 
$\Lambda_0$, $\Lambda_1$ is the same as that given in Lee \& Saio (1990).

The oscillation equations derived above are integrated in the thin surface magnetic layer, and
non-magnetic oscillation equations (e.g., Lee \& Saio 1986) are integrated in the non-magnetic core.
The two integrations are matched at the interface between the magnetic and non-magnetic regions to
obtain a complete solution.

For numerical computation, we solve a finite set of the linear ordinary differential equations, 
which is derived by
truncating the infinite expansions of perturbed quantities so that the first $k_{max}$ expansion
coefficients with $l_j$ and $l^\prime_j$ from $j=1$ to $j=k_{max}$ are retained for each quantity.
In most cases, we employ $k_{max}=6$ in this paper.
Using a relaxation method (e.g., Unno et al 1989),
we solve as an eigenvalue problem for $\bar\omega$
the finite set of linear differential equations
with appropriate boundary conditions at the center and the surface of the star and jump conditions
at the interface (see the next subsection).

\subsection{Bounday conditions and jump conditions}

The surface boundary conditions we use are 
\begin{equation}
\delta p/p=0, \quad i\bmath{b}^T=0, \quad \bmath{b}^S+L^+\bmath{b}^H=0,
\end{equation}
where $\delta p$ is the Lagrangian perturbation of the pressure, and $(L^+)_{ij}=\delta_{ij}(l^\prime_j+1)$.
The latter two conditions come from the assumption that 
the perturbed magnetic field must be continuous to the outside field that is 
given by a scalar potential
(i.e., $\nabla\times \bmath{B}^\prime=0$;
Cambell \& Papaloizou 1986, Cunha \& Gough 2000, Saio \& Gautschy 2004).
The inner boundary condition at the stellar center
is the regularity condition of the functions $\bmath{y}_1$
and $\bmath{y}_2$ (e.g., Lee \& Saio 1986).

The jump conditions we apply at the interface are the continuity of the 
functions $\bmath{y}_1$ and $\bmath{y}_2$, and the condition
that the oscillation energy imparted by the perturbed
magnetic field is lost from the wave system.
To consider the latter condition, it is instructive to
differentiate equation (18) with respect to $\ln r$ to obtain
\begin{equation}
r{d\over dr}r{d\over dr}\left(\matrix{\bmath{b}^H/\alpha \cr i\bmath{b}^T/\alpha\cr}\right)
=-{\alpha\over 2}\left(\matrix{-G_1^{-1}M_0\Lambda_0 &
mG_1^{-1}-\nu\bmath{1}\cr mG_0^{-1}-\nu\bmath{1}&-G_0^{-1}M_1\Lambda_1\cr}\right)
\left(\matrix{-G_0^{-1}M_1\Lambda_1&mG_0^{-1}\cr
mG_1^{-1}&-G_1^{-1}M_0\Lambda_0\cr}\right)
\left(\matrix{\bmath{b}^H/\alpha \cr i\bmath{b}^T/\alpha\cr}\right)+\cdots
\end{equation}
where we have used equations (16) $\sim$ (20) to eliminate $\bmath{h}$ and $i\bmath{t}$.
Since the quantity $|\alpha|$ can be very large in the interior,
if we assume a WKBJ type solution in the region near the interface such that
$(\bmath{b}^H/\alpha , i\bmath{b}^T/\alpha)\propto\exp\left(ik_r\ln r+im\phi+i\omega t\right)$,
we obtain the wavenumber $k_r\propto \sqrt{\alpha}$, which suggests the
existence of very short magnetic perturbations.
These short magnetic perturbation are attributable to the Alfv\'en waves. 
In the deep interior where
the gas pressure dominates the magnetic pressure so that $\beta>>1$, it may be legitimate to write
\begin{equation}
\left(\matrix{\bmath{b}^H/\alpha \cr i\bmath{b}^T/\alpha\cr}\right)
=\left(\matrix{\bmath{b}^H/\alpha \cr i\bmath{b}^T/\alpha\cr}\right)_0
+\left(\matrix{\bmath{b}^H/\alpha \cr i\bmath{b}^T/\alpha\cr}\right)_m, \quad
\left(\matrix{\bmath{h}\cr i\bmath{t}\cr}\right)=
\left(\matrix{\bmath{h}\cr i\bmath{t}\cr}\right)_0
+\left(\matrix{\bmath{h}\cr i\bmath{t}\cr}\right)_m, \quad
\left(\matrix{\bmath{y}_1\cr\bmath{y}_2\cr}\right)=
\left(\matrix{\bmath{y}_1\cr\bmath{y}_2\cr}\right)_0,
\end{equation}
where the functions with a subscript $0$ denote non-Alfv\'enic perturbations 
and those with a subscript $m$ denote the Alfv\'enic perturbations, and the former are
assumed to have wavelengths much longer than the latter.
For the non-Alfv\'enic perturbations in the deep interior, we assume 
\begin{equation}
\left(\matrix{-G_1^{-1}M_0\Lambda_0 &
mG_1^{-1}-\nu\bmath{1}\cr mG_0^{-1}-\nu\bmath{1}&-G_0^{-1}M_1\Lambda_1\cr}\right)
\left(\matrix{\bmath{h}\cr i\bmath{t}\cr}\right)_0=
-\left(\matrix{m\nu G_1^{-1}(K+M_0) & G_1^{-1}M_0\Lambda_0/c_1\bar\omega^2\cr
-\nu G_0^{-1}(m^2\Lambda_0^{-1}+M_1\Lambda_1K)&-mG_0^{-1}/c_1\bar\omega^2\cr}\right)
\left(\matrix{\bmath{y}_1\cr\bmath{y}_2\cr}\right)_0,
\end{equation}
which is equivalent to equations (A4) and (A5) in Lee \& Saio (1990), and
\begin{eqnarray}
&&{1\over 2}\left(\matrix{G_0^{-1}M_1\Lambda_1&-mG_0^{-1}\cr
-mG_1^{-1}&G_1^{-1}M_0\Lambda_0\cr}\right)
\left(\matrix{\bmath{b}^H \cr i\bmath{b}^T\cr}\right)_0=
r{d\over dr}\left(\matrix{\bmath{h}\cr i\bmath{t}\cr}\right)_0
-\left(\matrix{\bmath{1}+G_0^{-1}M_1\Lambda_1C_1/2&-mG_0^{-1}C_0/2\cr
-mG_1^{-1}C_1/2&\bmath{1}+G_1^{-1}M_0\Lambda_0C_0/2\cr}\right)
\left(\matrix{\bmath{h}\cr i\bmath{t}\cr}\right)_0\nonumber \\ 
&&-{1\over2}\left(\matrix{(V_G-4)G_0^{-1}(m^2\Lambda_0^{-1}+M_1\Lambda_1K) &
-V_GG_0^{-1}(m^2\Lambda_0^{-1}+M_1\Lambda_1K)\cr -m(V_G-4)G_1^{-1}(K+M_0)&
mV_GG_1^{-1}(K+M_0)\cr}\right)\left(\matrix{\bmath{y}_1\cr\bmath{y}_2\cr}\right)_0,
\end{eqnarray}
which comes from equation (4) assuming $\beta>>1$ (see \S 4).
Using equations (30) to (32) in equations (18) and (19), we obtain for the Alfv\'enic perturbations
\begin{eqnarray}
&&r{d\over dr}\left(\matrix{\bmath{b}^H/\alpha \cr i\bmath{b}^T/\alpha\cr}\right)_m
=\left(\matrix{f\bmath{1}&-mG_1^{-1}(C_1+M_0\Lambda_0)/2\cr
0&f\bmath{1}+G_0^{-1}(m^2\bmath{1}+M_1\Lambda_1C_1)/2\cr}\right)
\left(\matrix{\bmath{b}^H/\alpha \cr i\bmath{b}^T/\alpha\cr}\right)_m
+\left(\matrix{-G_1^{-1}M_0\Lambda_0 &
mG_1^{-1}-\nu\bmath{1}\cr mG_0^{-1}-\nu\bmath{1}&-G_0^{-1}M_1\Lambda_1\cr}\right)
\left(\matrix{\bmath{h}\cr i\bmath{t}\cr}\right)_m\nonumber \\ 
&&+\left(\matrix{f\bmath{1}&-mG_1^{-1}(C_1+M_0\Lambda_0)/2\cr
0&f\bmath{1}+G_0^{-1}(m^2\bmath{1}+M_1\Lambda_1C_1)/2\cr}\right)
\left(\matrix{\bmath{b}^H/\alpha \cr i\bmath{b}^T/\alpha\cr}\right)_0
-r{d\over dr}\left(\matrix{\bmath{b}^H/\alpha \cr i\bmath{b}^T/\alpha\cr}\right)_0
+\left(\matrix{\bmath{b}^S/\alpha \cr 0\cr}\right)_0,
\end{eqnarray}
\begin{eqnarray}
&&r{d\over dr}\left(\matrix{\bmath{h}\cr i\bmath{t}\cr}\right)_m=
{\alpha\over 2}\left(\matrix{G_0^{-1}M_1\Lambda_1&-mG_0^{-1}\cr
-mG_1^{-1}&G_1^{-1}M_0\Lambda_0\cr}\right)
\left(\matrix{\bmath{b}^H/\alpha \cr i\bmath{b}^T/\alpha\cr}\right)_m
+\left(\matrix{\bmath{1}+G_0^{-1}M_1\Lambda_1C_1/2&-mG_0^{-1}C_0/2\cr
-mG_1^{-1}C_1/2&\bmath{1}+G_1^{-1}M_0\Lambda_0C_0/2\cr}\right)
\left(\matrix{\bmath{h}\cr i\bmath{t}\cr}\right)_m.
\end{eqnarray}
Defining
\begin{equation}
\bmath{Y}_m=\left(\matrix{\bmath{b}^H/\alpha \cr \bmath{b}^T/\alpha \cr \bmath{h} \cr i\bmath{t}}\right)_m, 
\quad\quad
\bmath{y}_0=\left(\matrix{\bmath{y}_1\cr\bmath{y}_2\cr}\right)_0,
\end{equation}
we may formally rewrite equations (33) and (34) as
\begin{equation}
r{d\over dr}\bmath{Y}_m=A\bmath{Y}_m+B\bmath{y}_0,
\end{equation}
where equations (31) and (32) have been used to eliminate $(\bmath{b}^H_0,i\bmath{b}^T_0)$
and $(\bmath{h}_0,i\bmath{t}_0)$,
and matrices $A$ and $B$ are determined by equations (33) and (34).
Here, we regard equation (36) as a linear differential equation of $\bmath{Y}_m$ with
an inhomogeneous term $B\bmath{y}_0$. 
The homogeneous part of the linear differential equation
is solved by assuming $\bmath{Y}_m\propto e^{ik_r \ln r}$, where
the wavenumber $k_r$ is obtained 
as eigenvalues to the equation $ik_r\bmath{Y}_m=A\bmath{Y}_m $.

To make the discussion more concrete, let us now consider
a finite set of the linear differential equations, which is
obtained by truncating the infinite expansions of perturbed quantities, keeping the first
$k_{max}$ expansion coefficients for each perturbed quantity.
In this case, the dimension of the vector $\bmath{Y}_m$ is $4 k_{max}$, and the 
number of eigenvalues $k_r$ to $ik_r\bmath{Y}_m=A\bmath{Y}_m $ is also $4 k_{max}$.
To construct jump conditions at the interface, we pick up $N=2k_{max}$ 
eigenvalues $k_r^j$ and the eigenfunctions $\bmath{Y}_m^j$ that satisfy the condition
\begin{equation}
{\rm Re}(k_r^j)<0 \quad\quad {\rm and} \quad\quad |{\rm Re}(k_r^j)|>|{\rm Im}(k_r^j)|,
\end{equation} 
or
\begin{equation}
{\rm Im}(k_r^j)<0 \quad\quad {\rm and} \quad\quad |{\rm Im}(k_r^j)|>|{\rm Re}(k_r^j)|,
\end{equation} 
and write a general solution to equation (36) as
\begin{equation}
\bmath{Y}_m=\sum_{j=1}^{N} c_j\bmath{Y}_m^je^{ik_r^j\ln r}-A^{-1}B\bmath{y}_0
=\left(\bmath{Y}_m^1, \cdots, \bmath{Y}_m^N\right)\left(\matrix{c^1e^{k_r^1\ln r}\cr
\vdots\cr c^Ne^{k_r^N\ln r}\cr}\right)-A^{-1}B\bmath{y}_0
\end{equation}
where $-A^{-1}B\bmath{y}_0$ is a special solution to equation (36), and $c^j$'s are
arbitrary constants.
The condition (37) means that the Alfv\'enic perturbations are chosen to be running waves into the 
core (e.g., Saio \& Gautschy 2004).
It is important to note that, different from the case of acoustic modes, 
the number of the eigenvalues $k_r^j$ that satisfy
${\rm Re}(k_r)<0$ and $|{\rm Re}(k_r)|>|{\rm Im}(k_r)|$ is usually less than $N=2k_{max}$ 
for rotational modes.
This is the reason why we add the eigenvalues $k_r^j$ that satisfy
${\rm Im}(k_r)<0$ and $|{\rm Im}(k_r)|>|{\rm Re}(k_r)|$ to make the total number of the conditions
equal to $N$.
Splitting the solution (39) into the upper and the lower parts so that 
$\left(\bmath{Y}_m^1, \cdots, \bmath{Y}_m^N\right)_U$ and 
$\left(\bmath{Y}_m^1, \cdots, \bmath{Y}_m^N\right)_L$ become square martices of
dimension $2k_{max}\times 2k_{max}$, we have
\begin{equation}
\left(\bmath{Y}_m+A^{-1}B\bmath{y}_0\right)_U=
\left(\bmath{Y}_m^1, \cdots, \bmath{Y}_m^N\right)_U\left(\matrix{c^1e^{k_r^1\ln r}\cr
\vdots\cr c^Ne^{k_r^N\ln r}\cr}\right), 
\end{equation}
and
\begin{equation}
\left(\bmath{Y}_m+A^{-1}B\bmath{y}_0\right)_L=
\left(\bmath{Y}_m^1, \cdots, \bmath{Y}_m^N\right)_L\left(\matrix{c^1e^{k_r^1\ln r}\cr
\vdots\cr c^Ne^{k_r^N\ln r}\cr}\right).
\end{equation}
Eliminating the arbitrary constants $c^j$ between equations (40) and (41), we finally obtain
(e.g., Saio \& Gautschy 2004)
\begin{equation}
\left(\bmath{Y}_m^1, \cdots, \bmath{Y}_m^N\right)_U^{-1}\left(\bmath{Y}_m+A^{-1}B\bmath{y}_0\right)_U=
\left(\bmath{Y}_m^1, \cdots, \bmath{Y}_m^N\right)_L^{-1}\left(\bmath{Y}_m+A^{-1}B\bmath{y}_0\right)_L,
\end{equation}
where $\bmath{Y}_m$ is to be replaced by $\bmath{Y}-\bmath{Y}_0$ with
$\bmath{Y}_0=(\bmath{b}^H_0,i\bmath{b}^T_0,\bmath{h}_0,i\bmath{t}_0)$.
The number of jump conditions supplied by equation (42) is $N$, and since
the continuity of the functions $\bmath{y}_1$ and $\bmath{y}_2$ at the interface gives
$N$ jump conditions, we have in total 2$N$ conditions at the interface.

\section{Numerical results}

$R$-modes of non-magnetic polytropes were investigated, for example, 
by Yoshida \& Lee (2000a,b).
The Coriolis force is the restoring force, and
the oscillation frequency $\omega$ is proportional to the rotation frequency $\Omega$.
The $r$-modes are retrograde modes, in the corotating frame of the star, having the asymptotic frequency 
$\omega=2m\Omega/[l^\prime(l^\prime+1)]$ for $\Omega\rightarrow 0$ for given $m$ and $l^\prime$, and
the eigenfunctions are dominated by the toroidal component $iT_{l^\prime}$ of the 
displacement vector $\bmath{\xi}$.
For isentropic stars, the odd $r$-mode with $l^\prime=|m|$ is 
the only $r$-mode for a given $m$, and the eigenfunctions are
dominated by the nodeless toroidal component $iT_{|m|}$.

For neutron star models, we employ an $N=1$ polytrope of 
mass $M=1.4M_\odot$ and radius $R=10^6$cm, for which we assume $\Gamma_1=1+1/N$ so that
the Schwartzshild discriminant $A$ vanishes in the interior.
As discussed in the previous section, 
the radial wavenumber of Alfv\'enic perturbations near the interface may be approximately
given by 
\begin{equation}
|k_r|\sim \sqrt{\alpha/2}=t_A\omega/2\propto |\bar\omega|/B_S, 
\end{equation}
where $B_S=B_0(R)$, and $t_A=r/v_A$ with $v_A=\sqrt{B^2/4\pi\rho}$ being the Alfv\'en velocity
denotes the traveling time for the Alfv\'en waves.
This equation indicates that Alfv\'enic perturbations have very short wavelengths $\sim1/k_r$ 
if the traveling time of
the Alfv\'en waves is much longer than the typical oscillation period of the modes.
Figure 1 gives the quantity $\sqrt{\alpha}$ as a function of $x=r/R$ for $B_S=10^{12}$G (solid line),
$B_S=10^{14}$G (dashed line), and $B_S=10^{16}$G (dash-dotted line), assuming $\bar\omega=1$.
This figure shows that, as the depth $z\equiv R-r$ from the surface increases, 
the radial wavenumber $k_r$ of Alfv\'enic perturbations increases 
quite rapidly and becomes as large as $10^4\sim10^5$ beneath the surface, indicating that
the travelling time of the Alfv\'en waves in a typicl neutron star model
with $B_S\ltsim 10^{14}$G is much longer than the oscillation period of rotational modes
unless $\bar\omega\sim 0$.
Since the Alfv\'enic perturbations associated with the $r$- modes
have very short wavelengths for $B_S\ltsim 10^{14}$G, we have to allocate a large number 
of mesh points to describe them correctly even in the thin surface region.

\begin{table*}
\centering
\begin{minipage}{140mm}
\caption{Complex eigenvalue $\kappa=\bar\omega/\bar\Omega$ 
and the frequency deviation $(\kappa_R-\kappa_0)/\kappa_0$
of the $l^\prime=|m|$ $r$-modes at $\bar\Omega=0.1$.}
\begin{tabular}{@{}ccccccccc@{}}
\hline
 &\multicolumn{4}{c}{$m=1$} &  \multicolumn{4}{c}{$m=2$}\\
$B_S$(G)& $\kappa_R$ &$\kappa_I$& $(\kappa_R-\kappa_0)/\kappa_0$ & $x_i$ & 
       $\kappa_R$ &$\kappa_I$& $(\kappa_R-\kappa_0)/\kappa_0$ & $x_i$ \\
\hline
$10^{12}$& 0.99642&$2.7\times 10^{-13}$ &$1.5\times10^{-11}$ & 0.998 
         & 0.66517&$1.1\times 10^{-10}$ &$2.4\times10^{-9}$ & 0.997 \\
$10^{14}$& 0.99642&$ 5.0\times 10^{-9}$ &$-1.8\times10^{-8} $ & 0.98 
         & 0.66518&$ 9.5\times 10^{-7}$ &$1.8\times10^{-5}$ & 0.98 \\
\hline
\end{tabular}
\end{minipage}
\end{table*}

In Table 1, we tabulate the complex eigenvalue $\kappa\equiv\omega/\Omega=\kappa_R+i\kappa_I$ and 
the frequency deviation $(\kappa_R-\kappa_0)/\kappa_0$
for the $r$-modes of $m=1$ and $m=2$ at $\bar\Omega=0.1$
for $B_S=10^{12}$G and $B_S=10^{14}$G, where 
$\kappa_0$ denotes the eigenvalue of the non-magnetic $r$ modes, and
$x_{i}\equiv r_{i}/R$ and $r_i$ is the radial distance of the interface from the stellar center.
Here, we use $k_{max}=6$ for Table 1.
Note that the location $x_i$ of the interface for the case of $B_S=10^{14}$G is deeper than
that for $B_S=10^{12}$G.
We find that the eigenvalue $\kappa_R$ of the $r$-modes is hardly affected by
the magnetic field of strength $B_S\ltsim 10^{14}$G.
This is because the Alfv\'enic perturbations become
decoupled from the $r$-modes immediately beneath the surface, giving
only a skin effect on the modes.
If we consider a magnetic field as strong as $B_S\gtsim10^{16}$, the wavelengths 
of the Alfv\'enic perturbations
will be comparable to the radius of the star for rotational modes of $\bar\omega\sim\bar\Omega\ltsim1$, and
no decoupling of the Alfv\'enic perturbations from the rotational modes will be expected 
in a thin surface layer.
In this case, the method of calculation applied in this paper may not be justified, but
the linearized MHD equations need to be integrated
in the entire interior of the star to determine the eigensolutions.

Table 1 also shows that the $r$-modes are stable having positive $\kappa_I$, which is, however, quite small
for the strength of the field considered in this paper.
Although the magnetic effects on the eigenfrequency are weak, it is important to note
that the effects are larger for the larger $B_S$, and that
the $r$-mode of $m=1$ is less influenced by the dipole field than that of $m=2$,
as indicated by Table 1.

For the case of $B_S=10^{12}$G,
the real parts of the eigenfunctions of the $r$-mode of $m=1$
are plotted versus $r/R$ in Figures 2 to 4, and those of the $r$-mode of $m=2$ in Figures 5 to 7, 
where we assume $\bar\Omega=0.1$ and $k_{max}=6$, and we employ the amplitude normalization
given by $S_{l_1}=1$ at the surface.
The presence of the magnetic dipole field of strength $B_S=10^{12}$G
has practically no influence on the modal properties of the modes.
The magnetic $r$-mode has the dominant toroidal component $iT_{l^\prime=|m|}$ of the displacement
vector over the horizontal $H_{l=|m|+1}$ and radial $S_{l=|m|+1}$
components, and the amplitudes of the radial and horizontal components
are almost the same order of magnitude at the surface.
The Alfv\'enic perturbations $(\bmath{b}^H,i\bmath{b}^T)_m$ have very short wavelengths and small
amplitudes compared with those of $(\bmath{b}^H,i\bmath{b}^T)_0$.
The magnetic perturbations $(\bmath{h},i\bmath{t})_m$ themselves have very short wavelengths 
and very small amplitudes compared to those of $(\bmath{h},i\bmath{t})_0$.
We also find that $(\bmath{y}_1,\bmath{y}_2)$ is hardly affected 
by the magnetic perturbations near the surface.
It may be useful to note an approximate relation
$\left(\bmath{h}, i\bmath{t}\right)_m\sim\left(\bmath{b}^H,i\bmath{b}^T\right)_m/\sqrt{\alpha}$,
which is derived from equation (34) assuming $rd/dr\sim\sqrt{\alpha}$ 
for the Alfv\'enic perturbations.
Note that in the region near the interface, $(b_{l^\prime=1}^H)_0\sim$constant and $(ib_{l=2}^T)_0\sim0$ for the
$r$-mode of $m=1$, and $(b_{l^\prime=2}^H)_0\sim0$ and $(ib_{l=3}^T)_0\sim$constant for the $r$-mode of
$m=2$, where
$(b_{l^\prime}^H)_0$ and $(ib_{l}^T)_0$ denote the non-Alfv\'enic parts of 
$b_{l^\prime}^H$ and $ib_{l}^T$, respectively.
See \S 4 for a detailed discussion.

For the case of $B_S=10^{14}$G,
the real parts of the eigenfunctions of the $r$-mode of $m=2$
are plotted versus $r/R$ in Figures 8 to 10, 
where we assume $\bar\Omega=0.1$ and $k_{max}=6$, and we employ the amplitude normalization
given by $S_{l_1}=1$ at the surface.
The wavelengths and the amplitudes of the Alfv\'enic perturbations $(\bmath{b}^H,i\bmath{b}^T)_m$ are 
much longer and much larger than those found for the case of $B_S=10^{12}$G. 
It is interesting to note that the effect of the magnetic field on 
the eigenfunctions $\bmath{y}_1$ and $i\bmath{t}$ becomes apparent near the surface, as shown by
the inlets in Figures 8 and 9.
In Figure 11, to see the dependence of the Alfv\'enic perturbations on the value of $k_{max}$, we plot the
real parts of $(\bmath{b}^H,i\bmath{b}^T)$ for the $m=2$ $r$-mode at $\bar\Omega=0.1$ for $B_S=10^{14}$G, 
assuming $k_{max}=10$.
For this case, we have
the eigenvalue $(\kappa_R-\kappa_0)/\kappa_0=1.8\times10^{-5}$ and $\kappa_I=8.3\times10^{-7}$,
which is in good agreement with the value obtained for $k_{max}=6$.
Figure 11 shows that the wavelengths and the amplitudes of the Alfv\'enic perturbations 
$(\bmath{b}^H,i\bmath{b}^T)_m$ for $k_{max}=10$
becomes shorter and smaller than those for the case of $k_{max}=6$, which may reflect the property
of the wavenumber $k_r$ determined by $ik_r\bmath{Y}_m=A\bmath{Y}_m$
that $\max |{\rm Re}(k_r)|$ increases as $k_{max}$ increases.

\section{Discussions and Conclusions}

Let us rewrite equation (2) as
\begin{equation}
-c_1\bar\omega^2{\bmath{\xi}\over r}=-{r\nabla p^\prime\over \rho g r}-{\rho^\prime\over\rho}
-2ic_1\bar\omega\bar{\bmath{\Omega}}\times{\bmath{\xi}\over r}
+{2\over\beta V}{(r\nabla\times\bmath{B}^\prime)\times\bmath{B}\over B_0^2}.
\end{equation}
This equation may indicate that in the deep interior where the gas pressure dominates the magnetic pressure
such that $\beta>>1$, the magnetic perturbations $\bmath{B}^\prime$
have practically no influence on non-magnetic oscillations
governed by equation (44), and they are determined solely by the induction equation (4).
Using equation (4), we obtain
\begin{equation}
\nabla_H\cdot\bmath{B}^\prime_H=2B_0\left(r{\partial\over\partial r}-1\right)\left[-\cos\theta
\nabla_H\cdot{\bmath{\xi}_H\over r}+\sin\theta{\xi_\theta\over r}
+{1\over 2}\left(\sin\theta{\partial\over\partial\theta}+2\cos\theta\right){\xi_r\over r}\right],
\end{equation}
\begin{equation}
\left(\nabla_H\times \bmath{B}^\prime_H\right)_r=2B_0\left[
\left( r{\partial\over\partial r}-1\right)\left(-\cos\theta
\left(\nabla_H\times {\bmath{\xi}_H\over r}\right)_r-{im\over 2}{\xi_r\over r}\right)+\left(
r{\partial\over\partial r}
-1-{1\over 2}\nabla_H^2\right)\left(\sin\theta{\xi_\phi\over r}\right)\right],
\end{equation}
where
\begin{equation}
\bmath{B}^\prime_H=B^\prime_\theta\bmath{e}_\theta+B^\prime_\phi\bmath{e}_\phi, \quad
\bmath{\xi}_H=\xi_\theta\bmath{e}_\theta+\xi_\phi\bmath{e}_\phi, \quad
\nabla_H=\bmath{e}_\theta{\partial\over\partial\theta}+\bmath{e}_\phi{1\over\sin\theta}
{\partial\over\partial\phi}.
\end{equation}
Assuming 
\begin{equation}
\xi_r=0, \quad
\xi_\theta=T_{l^\prime}{1\over\sin\theta}{\partial\over\partial\phi}Y_{l^\prime}^m, \quad
\xi_\phi=-T_{l^\prime}{\partial\over\partial\theta}Y_{l^\prime}^m, 
\end{equation}
for the $r$-mode of a given $m$, we obtain from equations (45) and (46)
\begin{equation}
b^H_{l^\prime}=
-{2m\over l^\prime(l^\prime+1)} \left(r{\partial\over\partial r}-1\right){iT_{l^\prime}\over r},
\end{equation}
\begin{equation}
ib^T_l=
-{2l^\prime\over l^\prime+1}J^m_{l^\prime+1}
\left(r{\partial\over\partial r}+{l^\prime-1\over 2}\right){iT_{l^\prime}\over r},
\end{equation}
where $J_l^m=\sqrt{(l^2-m^2)/(4l^2-1)}$ and $l=l^\prime+1$ for odd modes.
Since $iT_{l^\prime}\propto r$ for the $r$-mode of $l^\prime=m=1$, we have
\begin{equation}
(b^H_{l^\prime})_0={iT_{l^\prime}\over r}={\rm constant}, 
\quad\quad (ib^T_l)_0=0,
\end{equation} 
and, since $iT_{l^\prime}\propto r^2$ for the $r$-mode of $l^\prime=m=2$, we have
\begin{equation}
(b^H_{l^\prime})_0=0, \quad\quad 
(ib^T_l)_0=-{2\over\sqrt{7}}{iT_{l^\prime}\over r}\propto r.
\end{equation}
Equations (51) and (52) for the non-Alfv\'enic parts of the magnetic perturbations
are consistent with the eigenfunctions shown in Figures 4 and 7.
Note that $(b^H_{l^\prime})_0$ and $(ib^T_l)_0$ are both non-zero for the $r$-modes of $l^\prime=m>2$.

Using a singular perturbation theory,
we calculate low-$m$ $r$-modes of a magnetized neutron star, 
assuming a magnetic dipole field whose axis is aligned with the axis of rotation.
For an $N=1$ polytrope of mass $M=1.4M_\odot$ and radius $R=10^{6}$cm,
we find that the magnetic field of strength $B_S\ltsim10^{14}$G has no significant influence 
on the modal properties of the $r$-modes with the frequency
$\bar\omega\sim\bar\Omega$ unless $\bar\Omega\sim 0$.
We show that the magnetic perturbations $\bmath{B}^\prime$ can be divided into Alfv\'enic and
non-Alfv\'enic parts, where the non-Alfv\'enic part in the deep interior is governed by
the induction equation (4).
The Alfv\'enic part, on the other hand, has very short wavelengths in the interior.
The radial wavenumber $k_r$ of the Alfv\'enic perturbations is approximately given by
$k_r\sim t_A\omega$ where $t_A$ denotes the traveling time of the Alfv\'en waves, and
we have $t_A\omega>>1$ beneath the stellar surface
for the $r$-modes we consider in this paper.
The Alfv\'enic perturbations become very short waves, being
decoupled from the $r$-modes in the interior.
This decoupling of the Alfv\'enic perturbations from the $r$-modes may cause a damping effect
on the modes.

If we consider the growthtime $\tau$ of the $r$-mode of $m=2$ driven unstable by current multipole 
gravitational radiation, we have
\begin{equation}
\tau\sim(3/4)^3\tilde{\tau}_{l=2}\bar\Omega^{-6}\sim1.4\bar\Omega^{-6} \quad {\rm sec},
\end{equation}
where we have used $\tilde{\tau}_{l=2}=3.31$sec, and $\bar\Omega=\Omega/\sigma_0$ with $\sigma_0=\sqrt{GM/R^3}$
(see, e.g, Yoshida \& Lee 2000).
Defining the growth timescale as $\tau=-1/\omega_I$, we have
\begin{equation}
\kappa_I\sim-5\times10^{-5}\bar\Omega^5,
\end{equation}
where we have used $\sigma_0=1.36\times10^4{\rm s}^{-1}$ for $M=1.4M_\odot$ and $R=10^6$cm.
Thus, we have $\kappa_I\sim -5\times10^{-10}$ for $\bar\Omega=0.1$ and
$\kappa_I\sim -1.2\times10^{-5}$ for $\bar\Omega=3/4$.
Comparing these values with the numbers $\kappa_I$ given in Table 1, we may find that
the $r$-mode instability can be damped by dissipative very short Alfv\'enic perturbations
in the surface layer when the neutron stars with $B_S>>10^{12}$G are slowly rotating.
For neutron stars in LMXB's, however, the strength of magnetic field is believed to be much weaker than
$10^{12}$G, and the effects of the magnetic field on the $r$-mode instability will be unimportant.

In this paper, we have assumed that the axis of the magnetic field is aligned with the axis of rotation.
If we lift this assumption, the problem will be much more complicated since the axisymmetry cannot be
assumed and hence a perturbed quantity must be
represented by a sum of terms proportional to $Y_l^m$ over the indices $l$ and $m$.
The possible existence of a crustal layer in the 
envelope of a rotating magnetized neutron star also makes difficult the problem of the oscillations
(e.g., McDermott, Van Horn, \& Hansen 1988; Lee \& Strohmayer 1996).
Magnetic effects on the torsional oscillations of a neutron star with a solid crust have been discussed by
Carroll et al (1986) and Messios, Papadopoulos, \& Stergioulas (2001), but
no effects of rotation have been included in these studies.
If there exists a fluid ocean on the solid crust of a rotating neutron star, the magnetic effects on
waves propagating in the ocean may be important observationally if the waves are 
responsible for burst oscillations found in neutron stars in LMXB systems (e.g., Lee 2004).

\vskip 0.5cm
\noindent{\sl  Acknowledgments:} 
{The author is very grateful to Prof. H. Saio for his illuminating discussion on the problem.}
%

%

\newpage

\begin{figure}
\centering
\epsfig{file=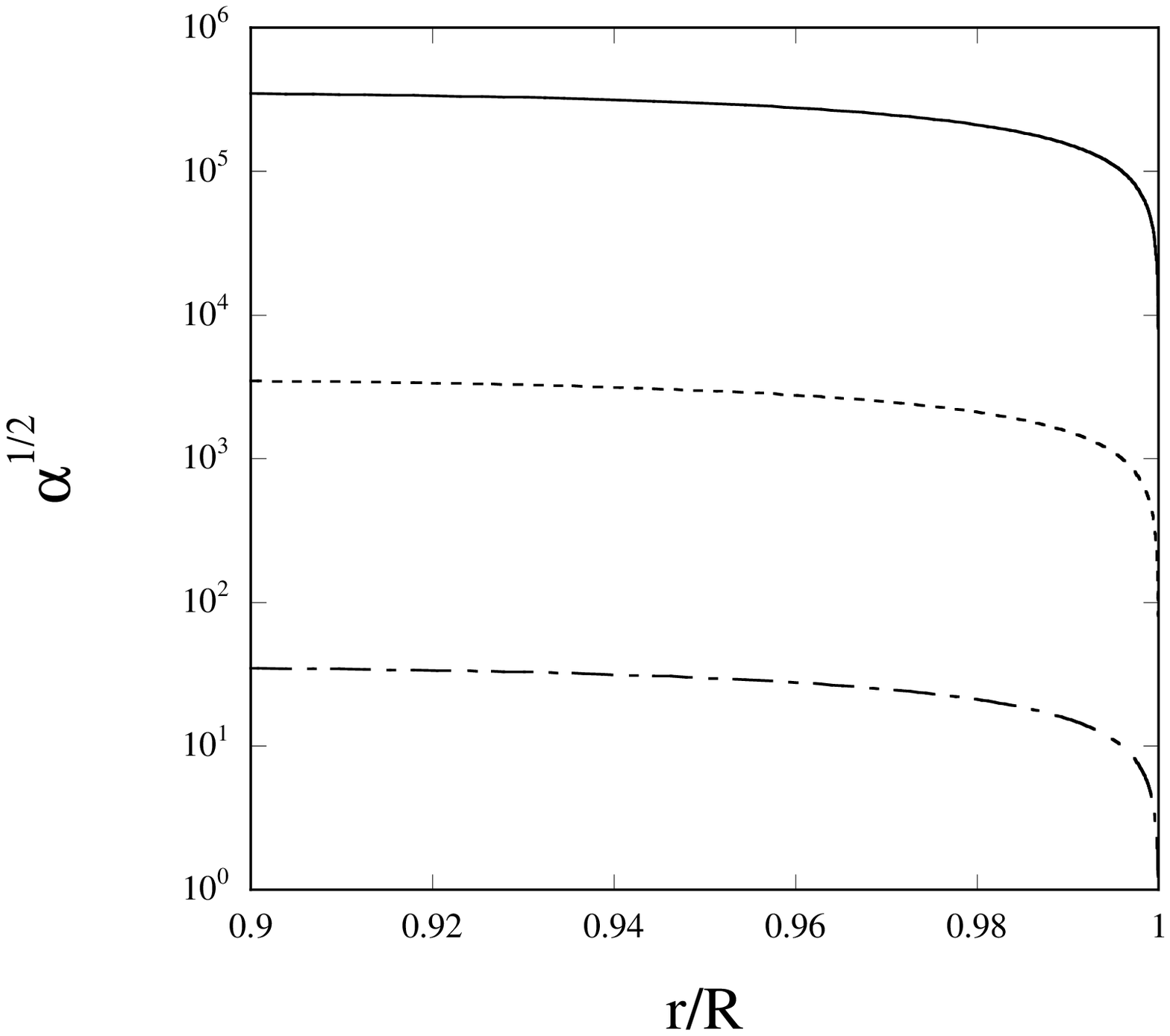,width=0.6\textwidth}
\caption{ Plot of $\alpha^{1/2}$ as a function of $r/R$ for an $N=1$
polytrope of mass $M=1.4M_\odot$ and radius $R=10^6$cm, where $\bar\omega=\omega/\sqrt{GM/R^3}=1$ 
is assumed. Solid, dashed, and dash-dotted lines are for the cases of $B_S=10^{12}$G, $10^{14}$G,
and $10^{16}$G, respectively.}
\end{figure}
\begin{figure}
\centering
\epsfig{file=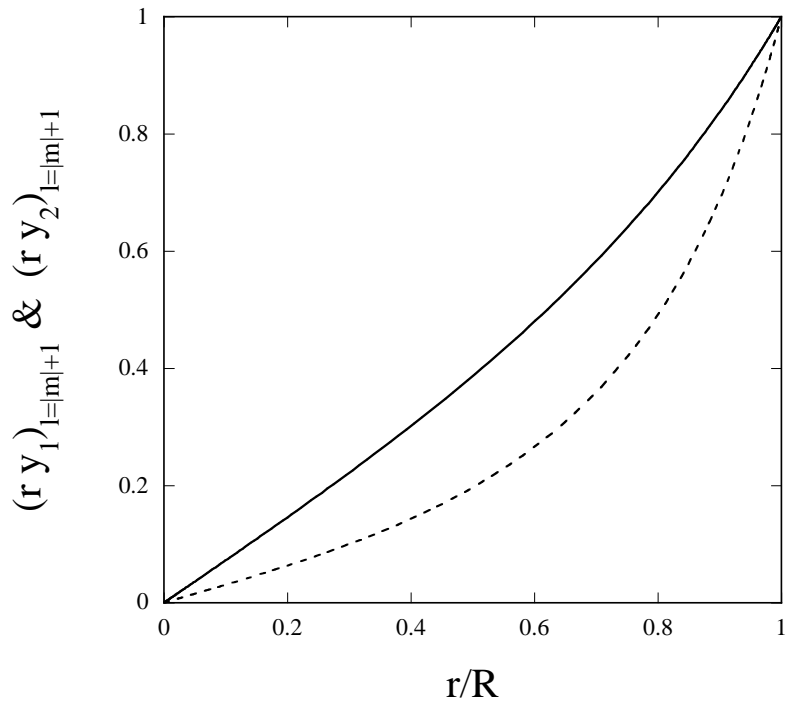,width=0.6\textwidth}
\caption{Eigenfunctions $(r\bmath{y}_1)_{l=|m|+1}$ (solid curve) and 
$(r\bmath{y}_2)_{l=|m|+1}$ (dashed curve)
as a function of $r/R$ for the $r$-mode of $m=1$ at $\bar\Omega=0.1$,
where $B_S=10^{12}$G and $k_{max}=6$.}
\end{figure}
\begin{figure}
\centering
\epsfig{file=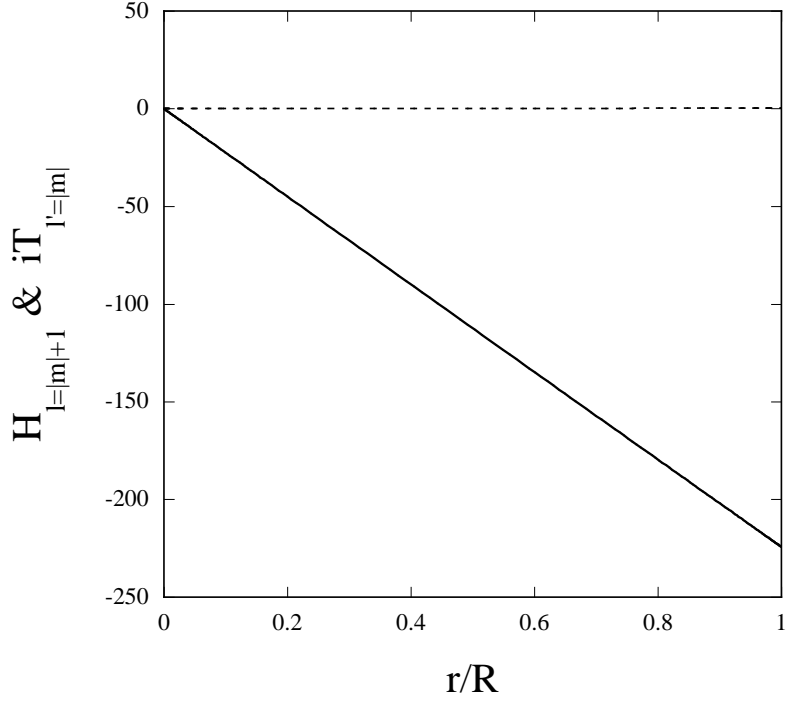,width=0.6\textwidth}
\caption{Eigenfunctions $iT_{l^\prime=|m|}$ (solid line)
and $H_{l=|m|+1}$ (dashed line) as a function of $r/R$ for the $r$-mode 
of $m=1$ at $\bar\Omega=0.1$, where $B_S=10^{12}$G and $k_{max}=6$.}
\end{figure}
\begin{figure}
\centering
\epsfig{file=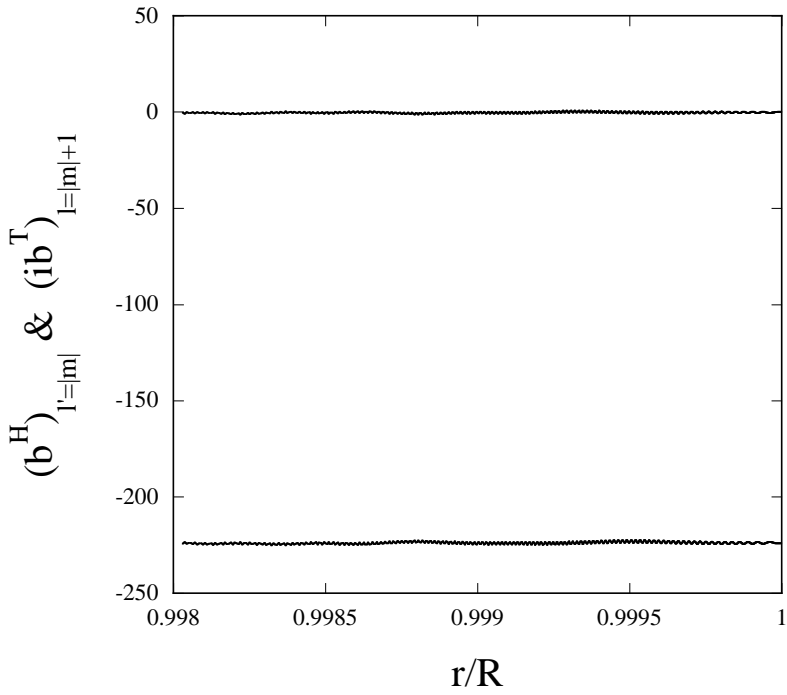,width=0.6\textwidth}
\caption{Eigenfunctions $b^H_{l^\prime=|m|}$ (lower curve)
and $ib^T_{l=|m|+1}$ (upper curve) as a function of $r/R$ for the $r$-mode 
of $m=1$ at $\bar\Omega=0.1$, where $B_S=10^{12}$G and $k_{max}=6$. }
\end{figure}
\begin{figure}
\centering
\epsfig{file=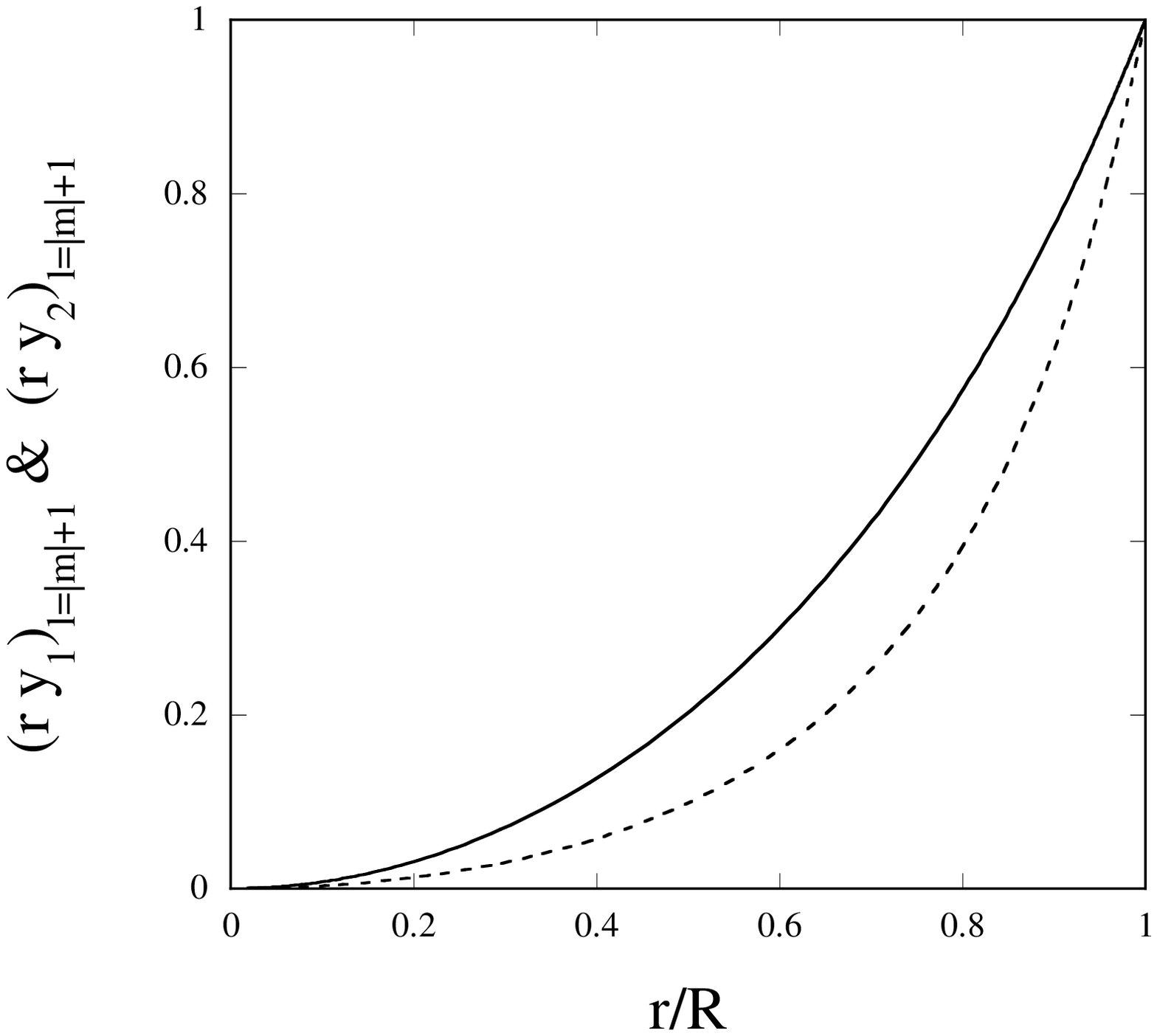,width=0.6\textwidth}
\caption{Same as Figure 2 but for $m=2$.}
\end{figure}
\begin{figure}
\centering
\epsfig{file=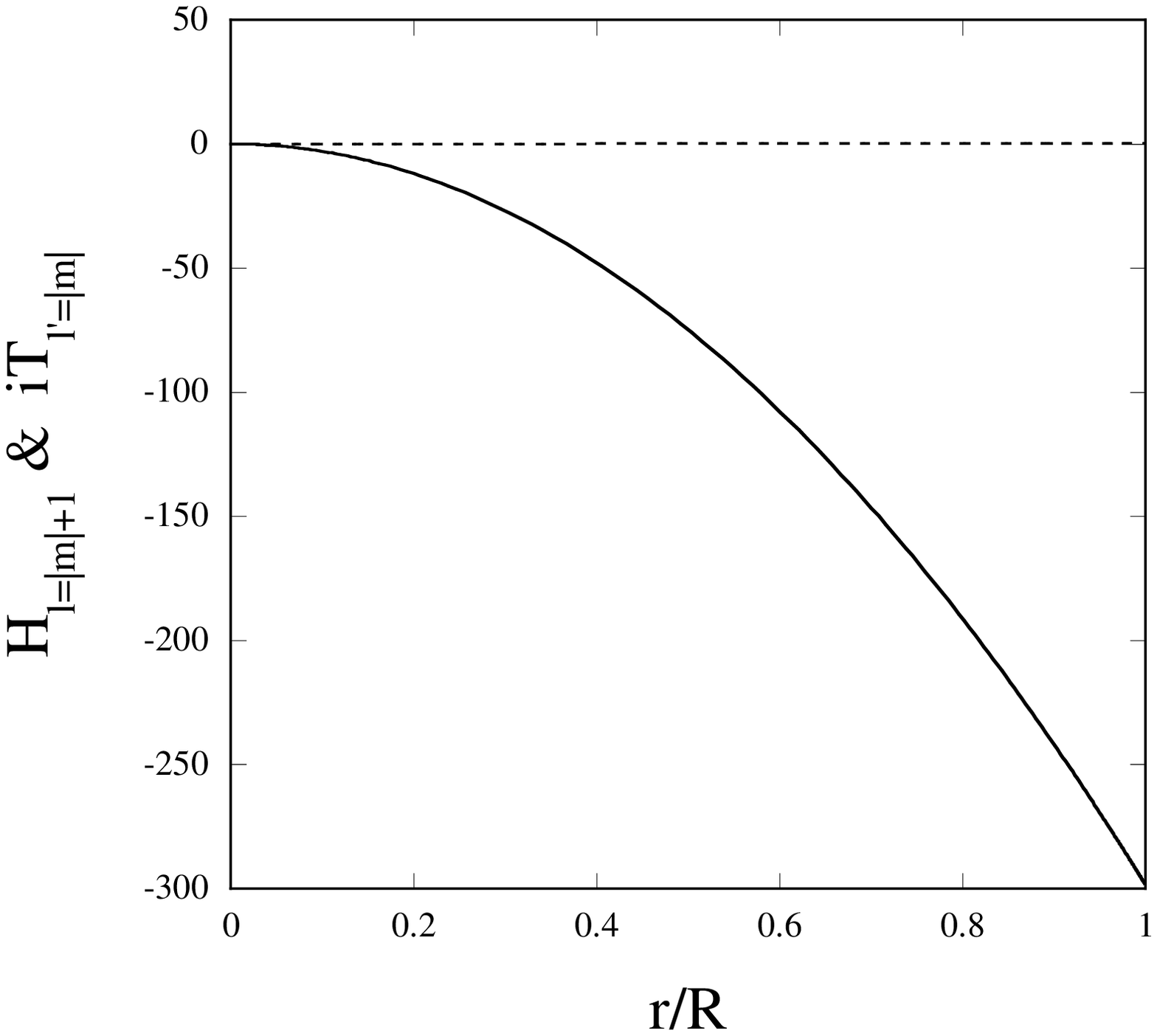,width=0.6\textwidth}
\caption{Same as Figure 3 but for $m=2$.}
\end{figure}
\begin{figure}
\centering
\epsfig{file=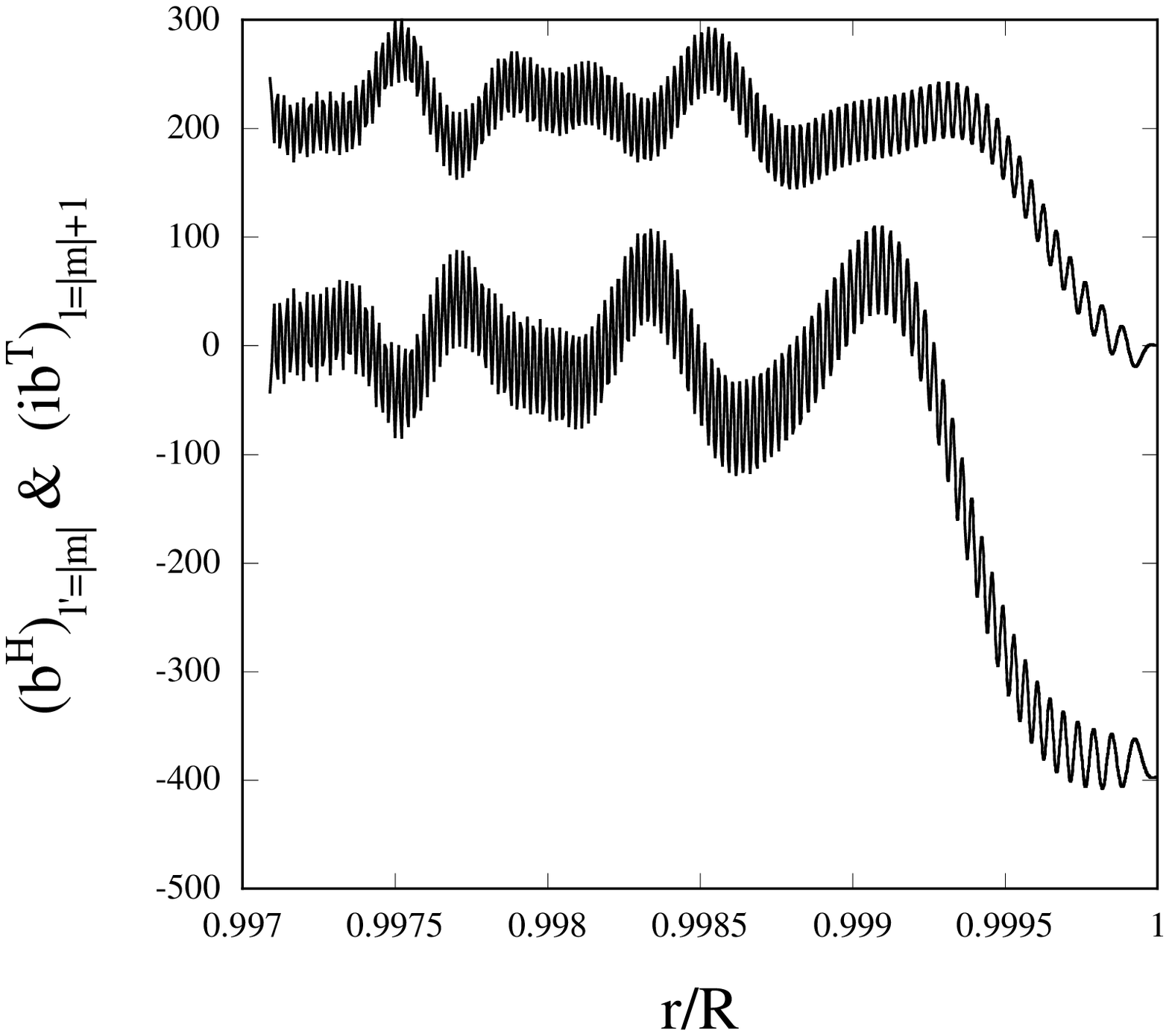,width=0.6\textwidth}
\caption{Same as Figure 4 but for $m=2$. }
\end{figure}
\begin{figure}
\centering
\epsfig{file=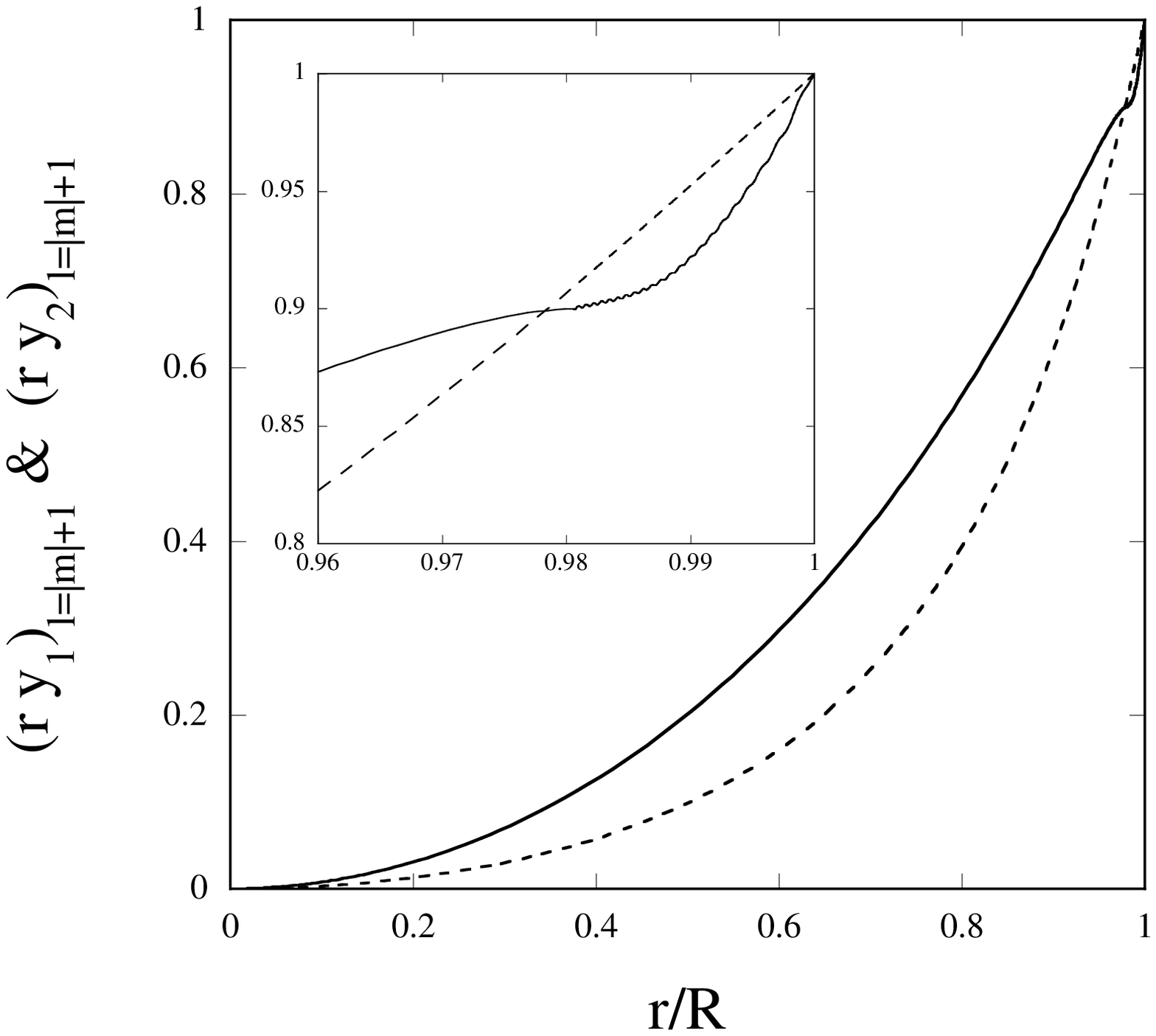,width=0.6\textwidth}
\caption{Same as Figure 5 but for $B_S=10^{14}$G. }
\end{figure}
\begin{figure}
\centering
\epsfig{file=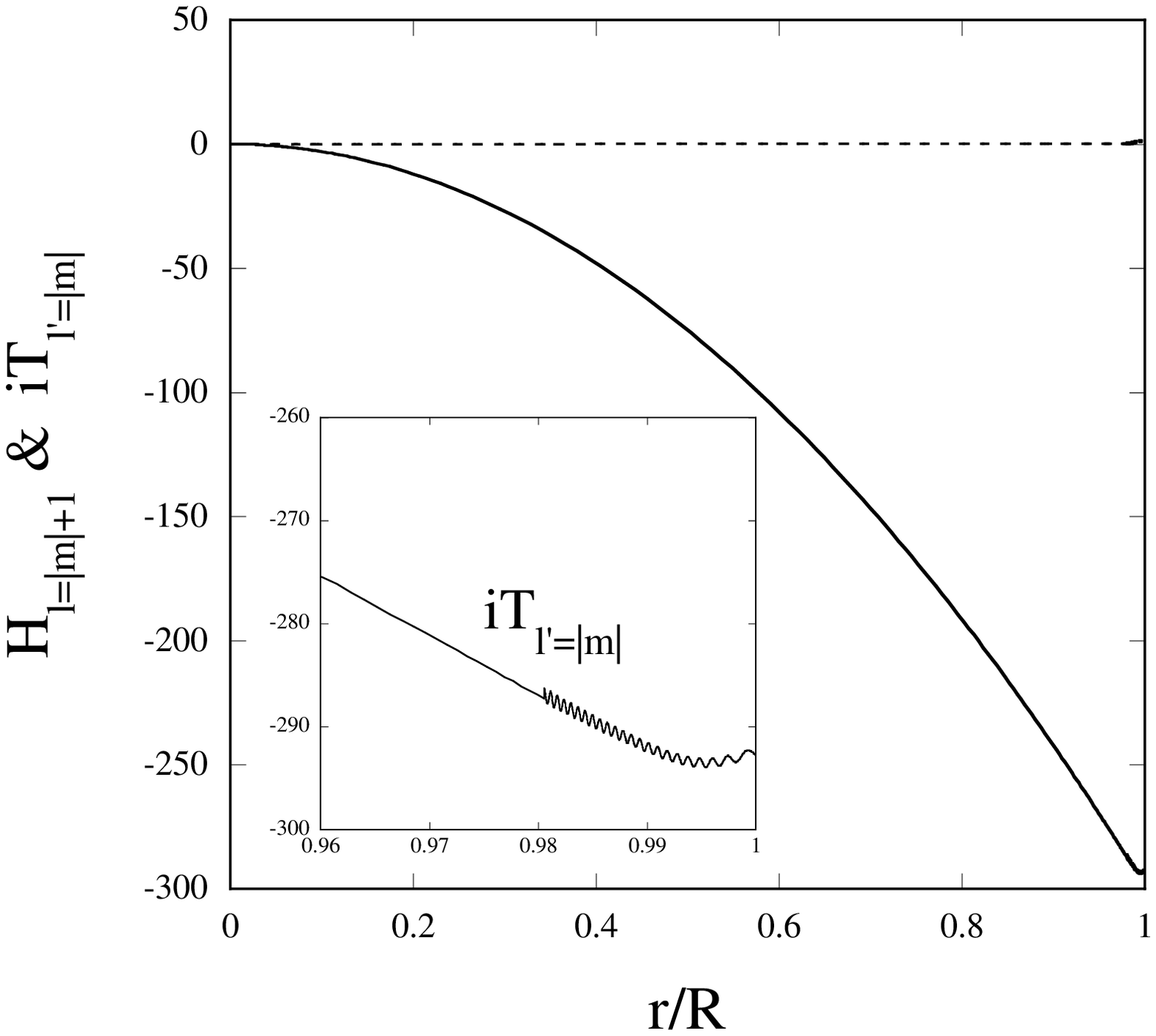,width=0.6\textwidth}
\caption{Same as Figure 6 but for $B_S=10^{14}$G. }
\end{figure}
\begin{figure}
\centering
\epsfig{file=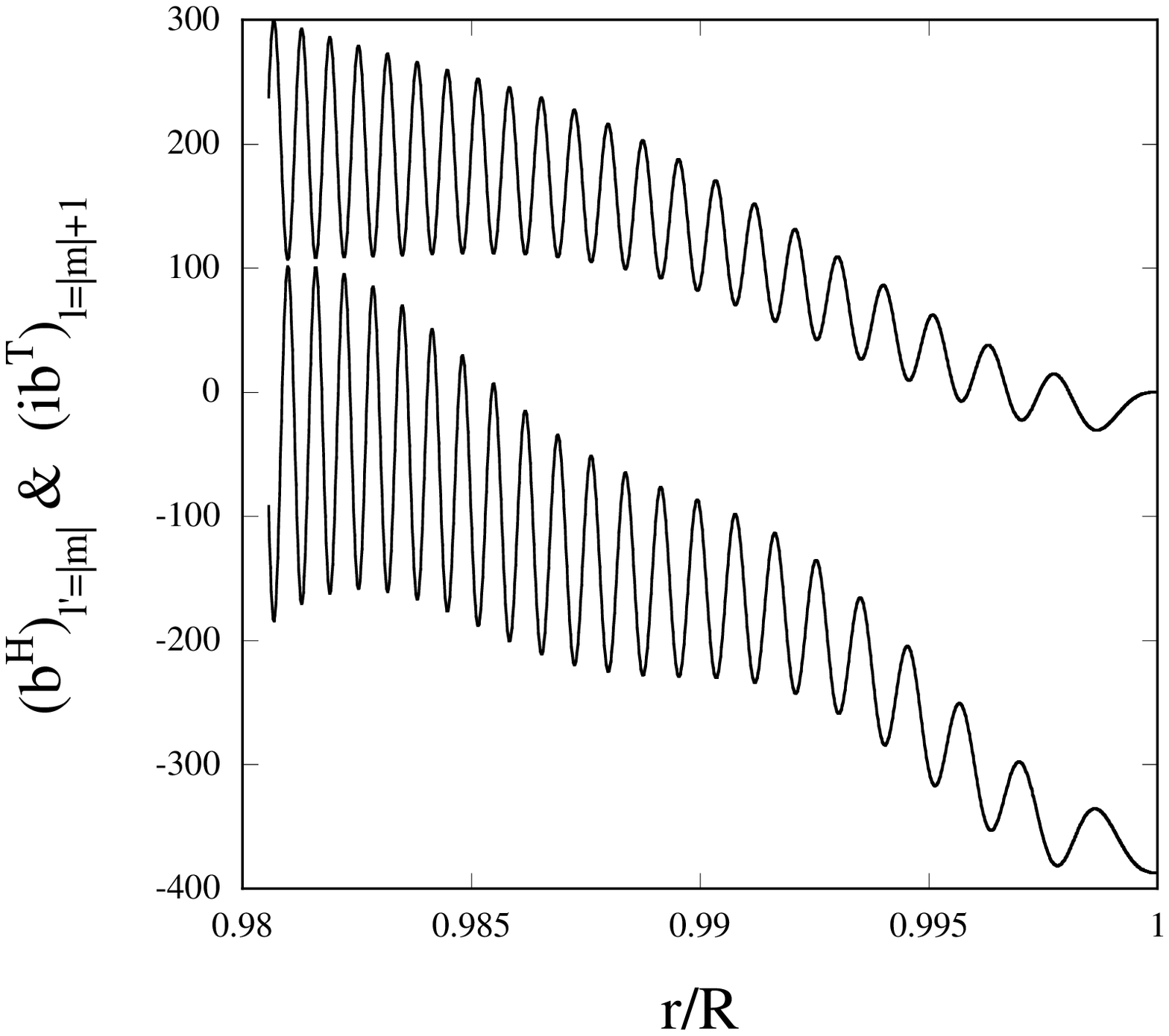,width=0.6\textwidth}
\caption{Same as Figure 7 but for $B_S=10^{14}$G. }
\end{figure}
\begin{figure}
\centering
\epsfig{file=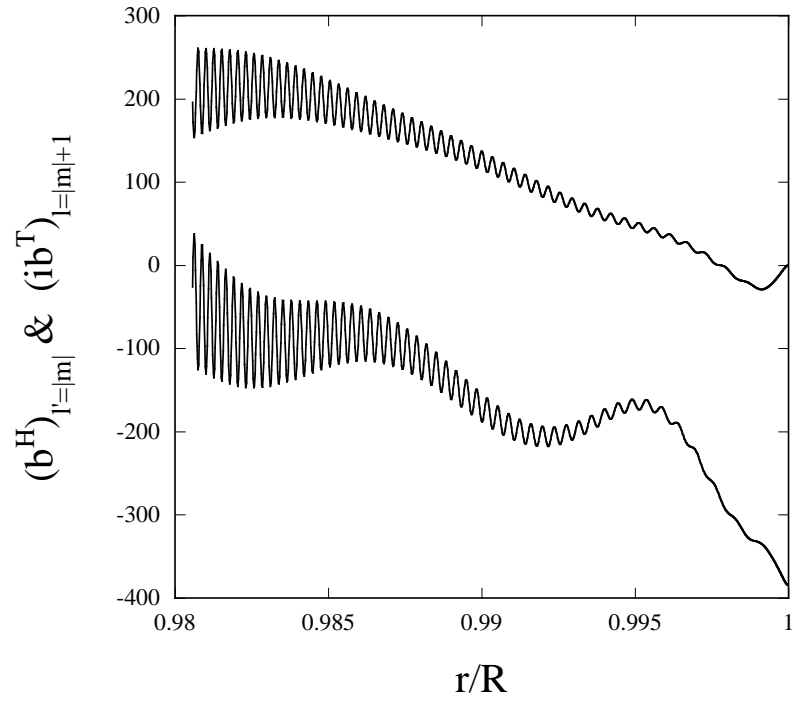,width=0.6\textwidth}
\caption{Same as Figure 10 but for $k_{max}=10$. }
\end{figure}

\end{document}